\documentclass[12pt,preprint]{aastex}

\newcommand{\teff}{$T_{\mathrm{eff}}$}
\newcommand{\teffs}{$T_{\mathrm{eff}} \;$}
\newcommand{\he}{HE 1305+0132}
\newcommand{\hes}{HE 1305+0132 $\;$}

\begin{document}

\title{FLUORINE IN A CARBON-ENHANCED METAL-POOR STAR}

\author{Simon C. Schuler\altaffilmark{1,2}, Katia Cunha\altaffilmark{1,3}, 
        Verne V. Smith\altaffilmark{1}, Thirupathi Sivarani\altaffilmark{4}, 
	Timothy C. Beers\altaffilmark{4}, AND Young Sun Lee\altaffilmark{4}}

\affil{
  \altaffiltext{1}{National Optical Astronomy Observatory/Cerro Tololo 
  Inter-American Observatory, Casilla 603, La Serena, Chile; 
  sschuler@ctio.noao.edu, cunha@noao.edu, vsmith@noao.edu}
  \altaffiltext{2}{Leo Goldberg Fellow}
  \altaffiltext{3}{on leave from Observat{\'o}rio Nacional, Rio de Janeiro,
  Brazil}
  \altaffiltext{4}{Department of Physics and Astronomy, Center for the Study of
  Cosmic Evolution, and Joint Institute for Nuclear Astrophysics, Michigan State
  University, East Lansing, MI; thirupathi@pa.msu.edu, beers@pa.msu.edu,
  leeyou25@msu.edu}
  }

\begin{abstract}
The fluorine abundance of the Carbon-Enhanced Metal-Poor (CEMP) star \hes has 
been derived by analysis of the molecular HF (1-0) R9 line at 2.3357 $\micron$ 
in a high-resolution ($R = 50,000$) spectrum obtained with the Phoenix 
spectrometer and Gemini-South telescope.  Our abundance analysis makes use of a
CNO-enhanced ATLAS12 model atmosphere characterized by a metallicity and CNO
enhancements determined utilizing medium-resolution ($R = 3,000$) optical and 
near-IR spectra.  The effective iron abundance is found to be $[\mathrm{Fe/H}] 
= -2.5$, making \hes the most Fe-deficient star, by more than an order of
magnitude, for which the abundance of fluorine has been measured.  Using 
spectral synthesis, we derive a super-solar fluorine abundance of 
$A(^{19}\mathrm{F}) = 4.96 \pm 0.21$, corresponding to a relative abundance of 
$[\mathrm{F/Fe}] = +2.90$.  A single line of the Phillips C$_2$ system is 
identified in our Phoenix spectrum, and along with multiple lines of the 
first-overtone vibration-rotation CO (3-1) band head, C and O abundances of 
$A(^{12}\mathrm{C}) = 8.57 \pm 0.11$ and $A(^{16}\mathrm{O}) = 7.04 \pm 0.14$ 
are derived.  We consider the striking fluorine overabundance in the framework 
of the nucleosynthetic processes thought to be responsible for the 
C-enhancement of CEMP stars and conclude that the atmosphere of \hes was 
polluted via mass transfer by a primary companion during its asymptotic giant 
branch phase.  This is the first study of fluorine in a CEMP star, and it 
demonstrates that this rare nuclide can be a key diagnostic of nucleosynthetic 
processes in the early Galaxy.
\end{abstract}

\keywords{stars:individual (HE 1305+0132) --- stars:abundances --- 
stars:atmospheres --- stars:carbon --- stars:population II}

\section{INTRODUCTION}
The single stable isotope of fluorine, $^{19}$F, does not contribute nor is 
synthesized in the main nuclear reactions taking place in the cores of stars, 
and its abundance in the Universe is the lowest of all the light elements ($6 
\leq Z \leq 20)$.  This and the notably limited number of atomic and molecular 
absorption lines in stellar spectra from which reliable abundances can be 
derived have made the nucleosynthetic origin of $^{19}$F the least understood 
of all the light elements \citep{clayton}.  Spectral features associated with 
the vibration-rotation transitions of the hydrogen fluoride molecule (HF) 
located in the near-IR K-band have proved to be the most accessible lines in 
stellar spectra, and the (1-0) R9 line at 2.3357 \micron, in particular, has 
been used in the analyses that have provided much of what is known about the 
nucleosynthesis and Galactic evolution of $^{19}$F (Jorissen, Smith, \& Lambert 
1992; Cunha et al. 2003; Cunha \& Smith 2005).

Based on theoretical calculations and scant observational data, three $^{19}$F
production sites have been identified.  \citet{1988Natur.334...45W} first 
suggested that it is produced in Type II supernovae (SNe) by neutrino 
spallation on $^{20}$Ne: $^{20}$Ne($\nu$,$\nu ' p$)$^{19}$F.  In a pair of 
companion papers, \citet{1992A&A...261..157F} and \citet{1992A&A...261..164J} 
provided evidence for $^{19}$F production during shell He-burning in asymptotic 
giant branch (AGB) stars from the reaction $^{14}$N($\alpha,\gamma$)$^{18}$
F($\beta^{+}$)$^{18}$O($p,\alpha$)$^{15}$N($\alpha,\gamma$)$^{19}$F; the source 
of the protons primarily being $^{14}$N($n,p$)$^{14}$C, with 
$^{13}$C($\alpha,n$)$^{16}$O providing the neutrons.  
\citet{2000A&A...355..176M} identified He-burning in Wolf-Rayet stars, where 
the same reactions as those in AGB stars could occur, as the third potential 
source.  The contribution of each site to the evolution of $^{19}$F in the 
Galaxy is not known, but the inclusion of all three sources in a Galactic 
chemical evolution model resulted in satisfactory agreement between the model 
prediction and the available observational data (Renda et al. 2004; Cunha \& 
Smith 2005).

Despite the lack of detailed knowledge of Galactic $^{19}$F production, 
observationally derived abundances of this rare nuclide provide valuable 
insight into the nucleosynthetic processes occurring in the Galaxy.  Herein we 
present the results of our $^{19}$F abundance analysis of the Carbon-Enhanced 
Metal-Poor (CEMP) star \object{HE 1305+0132} \citep{2001A&A...375..366C}.  As 
currently defined, CEMP stars are stars with $[\mathrm{C/Fe}] \geq +1.0$ 
\citep{2005ARA&A..43..531B}, and they fall into at least two general 
categories, those that exhibit large enhancements of s-process elements 
(CEMP-s) and those that do not have such enhancements (CEMP-no; Beers et al. 
2007a).  The origins of the chemical abundance patterns observed in CEMP-s and 
CEMP-no stars, respectively, are thought to be pollution by mass transfer from 
a primary companion during its AGB phase and formation from a cloud that was 
chemically enriched by a previous generation of massive stars (Ryan et al. 
2005; Aoki et al. 2007).  Low-metallicity stellar evolution models have 
demonstrated that the C overabundances, as well as those of N and O, observed 
in CEMP stars can be reproduced by both of the proposed sources (e.g., Lau, 
Stancliffe, \& Tout 2007; Hirschi 2007) and that the abundances of other 
elements, such as $^{19}$F, may be used to distinguish them (Meynet, 
Ekstr{\"o}m, \& Maeder 2006).  Indeed, \citet{2006A&A...447..623M} modeled a 
$7 \; M_{\sun}$ AGB star and a $60 \; M_{\sun}$ star, both with initial 
$[\mathrm{Fe/H}] = -3.6$ and initial rotational velocities of $v_{\mathrm{ini}} 
= 800 \mathrm{km} \; \mathrm{s}^{-1}$, and the AGB model produced approximately 
4 orders of magnitude more $^{19}$F than the massive star model.  This result 
implies that the derivation of $^{19}$F abundances in CEMP stars can place 
robust constraints on their chemical histories and that $^{19}$F may be a key 
diagnostic of nucleosynthetic processes at low metallicities.  In this paper we
describe the first investigation of $^{19}$F in a CEMP star.

\section{OBSERVATIONS AND ANALYSIS}
Single-order echelle K-band spectra (2.3305 - 2.3400 \micron) of \hes were 
obtained in 2007 March with the 8.1 m Gemini-South telescope using the NOAO 
Phoenix near-IR spectrometer in the $R = \lambda / \Delta \lambda = 50,000$ 
mode \citep{phoenix}.  A sequence of three integrations (1500 s each) was 
executed with the star positioned at three different locations separated by $4 
\arcsec$ along the slit.  Calibration frames, including darks, flats, and a 
spectrum of a hot rapidly rotating star used for correcting telluric 
contamination, were also obtained.  The raw 2-dimensional spectra were reduced 
to 1-dimensional spectra using routines in the IRAF software suite following 
the method described in \citet{2002AJ....124.3241S}.  The final co-added
spectrum has a per pixel signal-to-noise ratio of 129.

The procedure employed to derive the stellar parameters and overall metallicity 
of \hes is fully recounted in \citet{2007AJ....133.1193B} and is briefly 
described here.  The effective temperature (\teff) has been determined using 
published photometry and the color-\teffs calibrations of Alonso, Arribas, \& 
Martinez-Roger (1996).  We use the ($V-K$)-based \teffs estimate as it has been 
demonstrated to be the best photometric indicator of \teffs for CEMP stars 
(e.g., Cohen et al. 2002).  The $V$ magnitude ($V = 12.57$) is taken from 
\citet{2007ApJS..168..128B}, and the $K$ magnitude ($K = 9.814$) is from the 
Two Micron All Sky Survey Point Source Catalog \citep{2006AJ....131.1163S}.  
From these data we find $T_{\mathrm{eff}} = 4462 \pm 100 \; \mathrm{K}$.  The 
logarithm of the surface gravity ($\log g = 0.80 \pm 0.30$ in cgs units) has been estimated 
from the Padova isochrones \citep{2000A&AS..141..371G}, assuming a metallicity 
of $[\mathrm{m/H}] = -2.0$ and an age of 10 Gyr.  A microturbulence ($\xi$) of 
$2.00$ km $\mathrm{s}^{-1}$ has been assumed; the choice of $\xi$ is of little 
consequence here as the derived $^{19}$F abundance is not sensitive to its 
value.

The overall metallicity has been derived from medium-resolution optical and 
near-IR spectra of \hes obtained as part of a concerted program to estimate 
metallicities, CNO abundances, $^{12}$C/$^{13}$C ratios, and potential 
s-process element enhancement for as many CEMP stars as possible 
\citep{2007AJ....133.1193B}.  A preliminary model atmosphere characterized by 
the adopted stellar parameters, enhancements of CNO typical for CEMP stars, and 
otherwise subsolar metallicity was generated with the ATLAS12 stellar 
atmosphere code \citep{1996IAUS..176..523K}.  The model atmosphere was then 
used to produce synthetic spectra in the wavelength regions covered by our 
medium-resolution optical and near-IR spectra.  Comparisons of the synthetic 
and observed spectra provided new estimates of the metallicity and CNO 
enhancements, which were then used to generate a new model atmosphere.  This 
process was carried out until both the optical and near-IR spectra were 
simultaneously fit satisfactorily.  The final metallicity, CNO enhancements, 
and the associated uncertainties (as described in Beers et al. 2007a) are 
$[\mathrm{m/H}] = -2.50 \pm 0.50$, $[\mathrm{C/Fe}] = +2.20 \pm 0.35$, 
$[\mathrm{N/Fe}] = +1.60 \pm 0.46$, and $[\mathrm{O/Fe}] = +0.50 \pm 0.22$.

With the final ATLAS12 model characterized by the parameters described above and
the line list from \citet{2002AJ....124.3241S} and \citet{2003AJ....126.1305C}, 
we used the LTE stellar line analysis package MOOG \citep{1973ApJ...184..839S} 
to construct a synthetic spectrum of the 2.335 $\micron$ spectral region for 
comparison to the high-resolution Phoenix spectrum of \he.  In addition to the
HF line, there are features of particular interest to this and future analyses
of the 2.335 $\micron$ region in the spectra of CEMP stars.  These are a lone 
line at 2.3332 $\micron$ of the Phillips C$_{2}$ system and a handful of lines 
of the first-overtone vibration-rotation $^{12}$C$^{16}$O (3-1) band head.  
These C$_{2}$ and CO features can provide estimates of the stellar C and O 
abundances, respectively, and the C$_{2}$ line in particular is a serendipitous 
benefit of observing CEMP stars in this spectral region.  We have used the 
C$_{2}$ and CO features to derive the C and O abundances of \he; the fit to the 
C$_{2}$ line and one CO line are shown in Figure 1 and represent abundances of 
$A(^{12}\mathrm{C}) = \log N(^{12}\mathrm{C}) = 8.57 \pm 0.11$ and 
$A(\mathrm{^{16}\mathrm{O}}) = 7.04 \pm 0.14$.  In Figure 2, we show the 2.3357 
$\micron$ HF feature and the best synthetic fit, corresponding to an abundance 
of $A(\mathrm{^{19}\mathrm{F}}) = 4.96 \pm 0.21$.  Synthetic spectra 
representing $^{19}$F abundances $\pm 0.10$ dex of the best-fit abundance are 
also shown in Figure 2.  For each element, the quoted uncertainty is calculated 
by determining individually the sensitivities of the derived abundance to the 
adopted \teff, surface gravity, C enhancement, and metallicity, and then 
summing in quadrature the resulting uncertainties associated with each 
parameter.

\section{DISCUSSION}
As is evidenced by Figures 1 and 2, our synthetic spectrum fits well the
high-quality Phoenix spectrum of \he.  The relative C and O abundances derived 
from fitting the Phillips C$_{2}$ line and the lines of the CO (3-1) band head 
are $[\mathrm{C/Fe}] = +2.68$ and $[\mathrm{O/Fe}] = +0.88$, and in both cases,
the values are in agreement with those derived from the medium-resolution 
spectra within the combined uncertainties, which we note are considerable.  The 
synthetic fit to the HF line is particularly sensitive to the model \teffs and 
C-enhancement, with a $\pm 150 \; \mathrm{K}$ change in \teffs resulting in an 
abundance change of $\Delta A(^{19}\mathrm{F}) = \pm 0.48$ and a $\pm 0.50 \; 
\mathrm{dex}$ change in the model C-enhancement resulting in 
$\Delta A(^{19}\mathrm{F}) = \pm 0.31$; the sensitivity to \teffs and 
C-enhancement dominate the total uncertainty in the final $^{19}$F abundance.  
Raising the model \teffs and increasing the model opacity (increasing the 
C-enhancement) both effectively increase the temperature in the HF line-forming 
region, resulting in the higher derived $^{19}$F abundances.  The effect is 
also seen, although to a lesser degree, if the non-CNO metallicity of the model 
is increased, with a $\Delta [\mathrm{m/H}] = +1.0$ change resulting in a 
$\Delta A(^{19}\mathrm{F}) = +0.26$.

The $^{19}$F abundance of \hes is remarkable.  Adopting solar values from 
\citet{2005ASPC..336...25A}, $[\mathrm{m/H}] = -2.5$, and the O abundance 
derived from our high-resolution spectrum ($[\mathrm{O/H}] = -1.62$), the 
relative $^{19}$F abundances are $[\mathrm{F/Fe}] = +2.90$ and $[\mathrm{F/O}] 
= +2.02$.  \citet{2005ApJ...626..425C} and \citet{2003AJ....126.1305C} have 
derived $^{19}$F abundances of cool dwarfs in the Orion Nebula Cluster, as well
as reanalyzed the $^{19}$F abundances of K and M field giants from 
\citet{1992A&A...261..164J}, in order to investigate the evolution of $^{19}$F 
in the Galaxy.  \citet{2003AJ....126.1305C} found the $^{19}$F abundances of 
the field giants to fall nicely along the line of scaled solar abundance in the 
$A(^{19}\mathrm{F})$ versus $A(\mathrm{Fe})$ and $A(^{19}\mathrm{F})$ versus 
$A(^{16}\mathrm{O})$ planes, with scatter comparable to the analysis 
uncertainties.  \citet{2005ApJ...626..425C} combined the red giant data with 
new Orion Nebula Cluster data in the $\mathrm{[F/O]}$ versus 
$A(^{16}\mathrm{O})$ plane, along with a chemical evolution model from 
\citet{2004MNRAS.354..575R}.  The stars span less than 0.5 dex below the solar 
value in O abundance, but nonetheless the abundances of both stellar 
populations follow the chemical evolution model, which predicts a gradual 
decline in $\mathrm{[F/O]}$ with decreasing $A(^{16}\mathrm{O})$.  The large 
overabundance of $^{19}$F in \hes deviates significantly from the existing 
empirical data and the chemical evolution model of \citet{2004MNRAS.354..575R}, 
strongly suggesting that the origin of the $^{19}$F in \hes lies outside of the 
standard Galactic chemical evolution channel.

\hes is the most Fe- and O-poor star- {\it by at least 1.5 orders of 
magnitude}- for which the abundance of $^{19}$F has been derived, and while the
bulk metallicity of \he, as defined by elements such as Fe, Ca, or Ti, is very 
low ($[\mathrm{m/H}] = -2.5$), the $^{19}$F and $^{12}$C abundances are 
enhanced enormously relative to these other metals.  Numerous theoretical 
studies address the nucleosynthetic origins of the enhanced $^{12}$C that 
defines CEMP stars (e.g., Meynet et al. 2006; Hirschi 2007; Tominaga, Umeda, \& 
Nomoto 2007), with most efforts focused on two scenarios: mass transfer of 
processed material from a primary companion during its AGB phase and formation 
from material enriched by previous generation of massive stars.  The 
overabundance of $^{19}$F in \hes coupled to the overabundance of $^{12}$C 
places the nucleosynthetic origin of these abundance anomalies squarely in the 
realm of the AGB stars.  In Figure 3 the observed trend between $A(^{19}$F) and 
$A(^{12}$C) for MS, S, and C stars taken from \citet{1992A&A...261..164J}, as 
well as results for the hot He-stars from Werner, Rauch, \& Kruk (2005), are 
shown.  The Jorissen et al. sample contains both intrinsic thermally-pulsating
AGB (TP-AGB) stars and stars that have been polluted by an AGB companion, and 
it is seen that both stellar types follow the same overall trend of increasing 
$^{19}$F with increasing $^{12}$C.  The hot He-stars from 
\citet{2005A&A...433..641W} are essentially the exposed cores of former AGB 
stars and reveal directly the products of He-burning during the phase of TP-AGB 
evolution.  While there may be a systematic offset, or perhaps larger scatter, 
between the $^{19}$F--$^{12}$C trends as defined by the two studies, it must be
noted that the hot He-stars have $T_{\mathrm{eff}} = 90,000 - 200,000 \; 
\mathrm{K}$, and their $^{19}$F and $^{12}$C abundances are derived from highly
ionized (\ion{F}{6} and \ion{C}{3}) lines.  The MS, S, and C stars, on the 
other hand, have much lower temperatures ($T_{\mathrm{eff}} = 3000 - 4000 \; 
\mathrm{K}$) and have abundances derived from molecular lines of C$_{2}$, CO, 
and HF.  The similarity in the run of fluorine with carbon from these very 
different types of analyses provide strong evidence that one path for the 
nucleosynthesis of $^{19}$F is tied to the production of $^{12}$C during TP-AGB 
evolution.

The abundances of $^{19}$F and $^{12}$C measured here for \hes are also plotted 
in Figure 3 and fall right on the trend defined by the 
\citet{1992A&A...261..164J} sample of red giants.  Combined with the fact that 
the Jorissen et al. stars all have near-solar metallicities ($\sim$-0.5 to +0.5 
in [Fe/H]) while \hes has $\mathrm{[m/H]} = -2.5$, the position of \hes in
Figure 3 is evidence for efficient production of $^{19}$F in metal-poor AGB 
stars.  This efficiency is the result of two effects: the primary nature of the 
neutron source and the lack of heavy-metal neutron ``poisons'' in a 
low-metallicity environment.  The neutron producing reaction 
$^{13}$C($\alpha,n$)$^{16}$O is fueled by $^{13}$C synthesized via 
$^{12}$C(p,$\gamma$)$^{13}$N($\beta^{+}$,$\nu$)$^{13}$C by the mixing of
protons into primary $^{12}$C produced by the triple-$\alpha$ process, making 
the resulting neutrons primary (i.e., independent of metallicity).  The 
neutrons then take part in the proton producing reaction
$^{14}$N($n,p$)$^{14}$C, which then take part in the final reactions 
$^{18}$O(p,$\alpha$)$^{15}$N($\alpha$,$\gamma$)$^{19}$F.  Because at 
low-metallicity the heavy elements that could compete for neutron captures are 
largely absent, the neutrons are very efficient at driving the final reaction 
chain necessary for $^{19}$F production.

Indeed, \citet{2006A&A...447..623M} recently modeled a massive ($7 \; 
M_{\odot}$), rapidly rotating ($v_{\mathrm{ini}} = 800 \; \mathrm{km} \; 
\mathrm{s}^{-1}$), metal-poor ($[\mathrm{m/H}] = -3.6$) AGB star and computed the
chemical composition of the envelope at the beginning of the TP-AGB phase; 
$^{19}$F was found to be prodigiously produced, with a predicted envelope 
abundance of $[\mathrm{F/Fe}] \approx +4.0$.  While this predicted $^{19}$F 
abundance differs from the observed value by about an order of magnitude, 
theoretical yields of low-metallicity stellar models are highly dependent on 
initial metallicities, masses, and rotational velocities (Hirschi 2007;
Chiappini et al. 2006), and the mixing of
the accreted material in the secondary star is not well understood.  Thus, the near
agreement is encouraging and should provide sufficient motivation 
to expand efforts to model low-metallicity AGB stars.  Parenthetically, the 
near-agreement also suggests that rotation may have a significant impact on the 
nucleosynthetic yields of metal-poor stars.

The second hypothesis to account for the chemical composition of CEMP stars- 
formation from previously enriched material- must also be considered for \he. 
Models of the winds and SN ejecta of massive zero- and low-metallicity stars, 
like models of low-metallicity AGB stars, can effectively reproduce the 
enhanced CNO abundances observed in CEMP stars, but with regards to $^{19}$F, 
no prodigious overproduction is seen.  For example, \citet{2006A&A...447..623M} 
modeled a rapidly-rotating ($v_{\mathrm{ini}} = 800 \mathrm{km} \; 
\mathrm{s}^{-1}$) $60 \; M_{\sun}$ star at a metallicity of $[\mathrm{m/H}] = 
-3.6$ and found that the star loses $\sim 40\%$ of its initial mass via winds 
over its lifetime, and that this wind material has a solar-scaled $^{19}$F 
abundance ($[\mathrm{F/Fe}] \approx 0.0$).  \citet{2007ApJ...660..516T} 
performed hydrodynamic and nucleosynthesis core-collapse SN calculations of 
Population III ($Z = 0$) stars with main-sequence masses of 13 - 50 
$M_{\sun}$.  Their SN model yields are consistent with the abundance patterns 
of very metal-poor stars in general, but in all of the calculations, $^{19}$F 
is highly depleted ($[\mathrm{F/Fe}] < -2.0$).  The $^{19}$F production seen in
these two studies is typical of zero- and low-metallicity massive star 
nucleosynthesis models (e.g., Heger \& Woosley 2002), and they seem to be 
unable to account for the highly enhanced $^{19}$F abundance observed in \he.

The position of \he in the $A(^{19}$F) and $A(^{12}$C) plane shown in Figure 3 
soundly ties its nucleosynthetic history to that of TP-AGB stars and stars 
known to have been polluted by an AGB companion, and thus mass transfer from an 
AGB companion remains the preferred explanation for its observed abundance 
pattern.  Accordingly, the abundances of the s-process elements are predicted 
to be enhanced in \he.  \citet{2005MNRAS.359..531G} and 
\citet{2006Ap.....49..173G} provide circumstantial evidence for this as they 
have independently identified \hes as a CH star, which are known as a group to 
be enhanced in s-process elements \citep{1998ARA&A..36..369W}; confirmation of 
this expected enhancement awaits a high-resolution spectroscopic abundance 
analysis.  The derived large overabundances of $^{19}$F and $^{12}$C in \hes, 
relative to the near-solar metallicity AGB stars from 
\citet{1992A&A...261..164J} and \citet{2005A&A...433..641W}, are evidence that 
AGB star nucleosynthesis is highly efficient at low metallicities and point to 
the importance of AGB stars in the nucleosynthetic history of the early Galaxy.

\acknowledgements
Support for S.C.S. has been provided by the NOAO Leo Goldberg Fellowship; NOAO 
is operated by the Association of Universities for Research Astronomy, Inc., 
under a cooperative agreement with the National Science Foundation.  K.C. and
V.V.S. acknowledge support from the National Science Foundation under grant
AST 06-46790.  T.S., T.C.B., and Y.S.L. acknowledge partial support for this
work from the National Science Foundation under grants AST 04-06784, AST
07-07776, and PHY 02-16783; Physics Frontier Center/Joint Institute for Nuclear
Astrophysics (JINA).  This paper is based on observations obtained with the 
Phoenix infrared spectrograph, developed and operated by NOAO, and the Gemini 
Observatory, which is operated by the Association of Universities for Research 
in Astronomy, Inc., under a cooperative agreement with the NSF on behalf of the 
Gemini partnership: the National Science Foundation (United States), the 
Particle Physics and Astronomy Research Council (United Kingdom), the National 
Research Council (Canada), CONICYT (Chile), the Australian Research Council 
(Australia), CNPq (Brazil) and CONICET (Argentina).  The observations were 
conducted under Gemini program GS-2007A-DD-1.

{\it Facility: } \facility{Gemini:South (Phoenix)}, \facility{SOAR (OSIRIS)}


\clearpage


\begin{figure}
\plotone{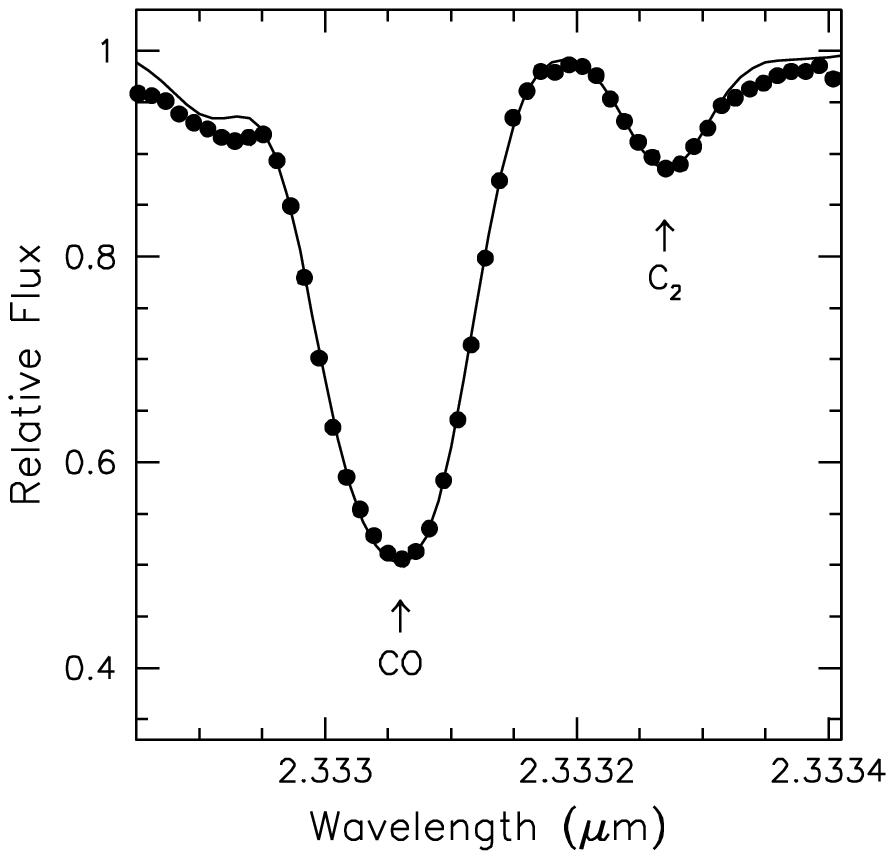}
\caption{High-resolution near-IR Phoenix spectrum of \hes (filled circles) and
the synthetic fit (solid line) of the $2.3331 \; \micron$ region.  The 
Phillips C$_{2}$ line at 2.3332 $\micron$ and a line of the first-overtone 
vibration-rotation $^{12}$C$^{16}$O (3-1) band head are marked.  This
Phillips line is the sole C$_{2}$ line in our Phoenix spectrum.}
\end{figure}

\begin{figure}
\plotone{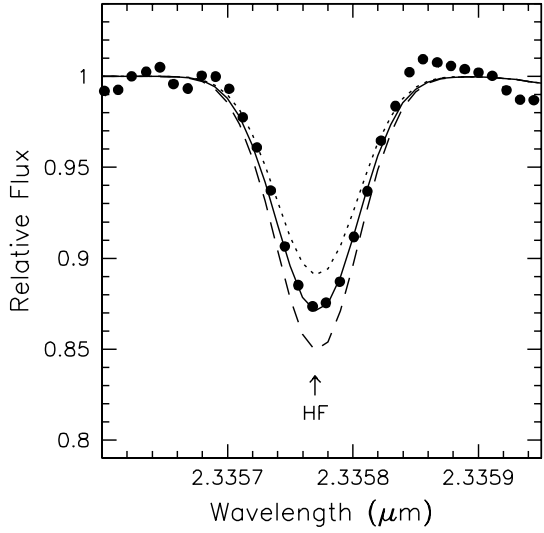}
\caption{HF (1-0) R9 line at 2.3357 $\micron$.  The observed high-resolution
near-IR Phoenix spectrum is shown as filled circles, and the best-fit synthetic 
spectrum, representing a derived $^{19}$F abundance of $A(\mathrm{^{19}F}) = 4.96 
\pm 0.21$, is given by the solid line.  The broken lines are synthetic spectra 
characterized by $^{19}$F abundances $\pm 0.10$ dex of the best-fit abundance.}
\end{figure}

\begin{figure}
\plotone{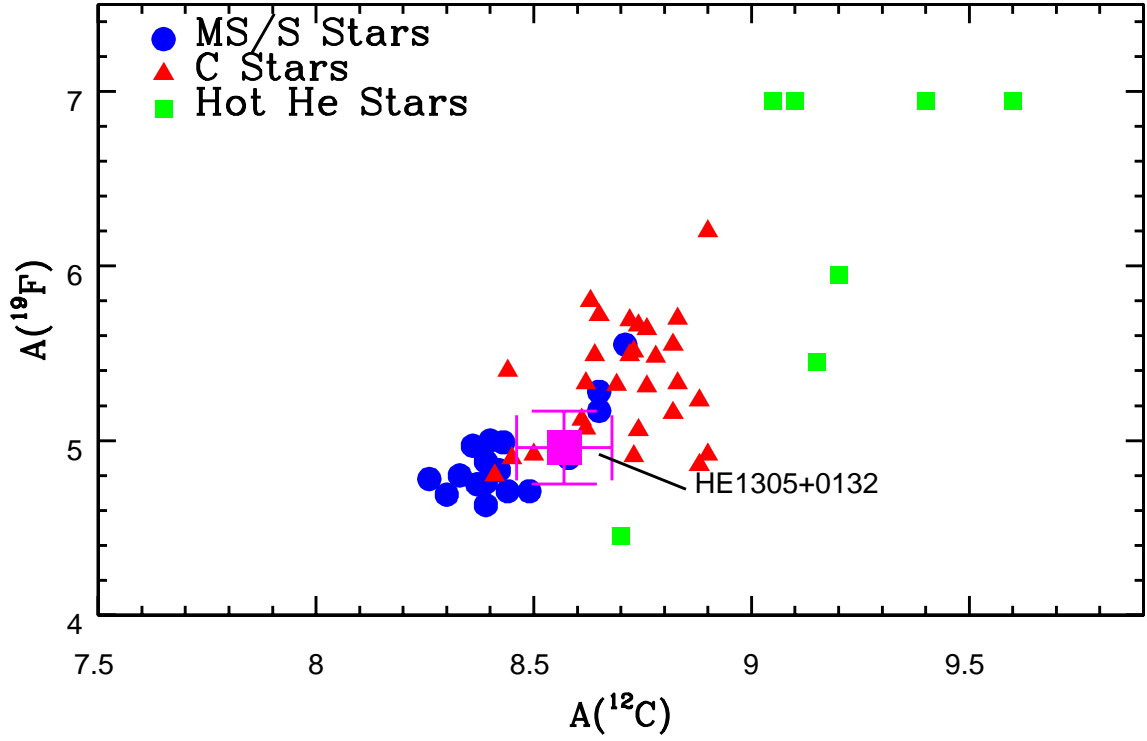}
\caption{Logarithmic abundances of $^{19}$F vs. $^{12}$C.  The abundances of 
\hes are marked by the magenta box with error bars.}
\end{figure}


\end{document}